\documentclass[twocolumn,british,pra,superscriptaddress,showpacs]{revtex4-1}
\usepackage[letterpaper]{geometry}
\usepackage[dvips]{graphicx}
\usepackage{amsmath}
\usepackage{amssymb}
\usepackage{esint}
\newcommand{\bra}[1]    {\langle #1|}
\newcommand{\ket}[1]    {| #1 \rangle}

\renewcommand{\t}[1]{\textrm{#1}}

\makeatletter
\@ifundefined{textcolor}{}
{%
 \definecolor{BLACK}{gray}{0}
 \definecolor{WHITE}{gray}{1}
 \definecolor{RED}{rgb}{1,0,0}
 \definecolor{GREEN}{rgb}{0,1,0}
 \definecolor{BLUE}{rgb}{0,0,1}
 \definecolor{CYAN}{cmyk}{1,0,0,0}
 \definecolor{MAGENTA}{cmyk}{0,1,0,0}
 \definecolor{YELLOW}{cmyk}{0,0,1,0}
 }

\makeatother

\makeatother

\usepackage{babel}

\begin{document}

\title{Phase estimation without a priori knowledge in the presence of loss
}

\author{Jan Ko{\l}ody{\'n}ski}
\affiliation{Institute of Theoretical Physics, University of Warsaw, ul. Ho\.{z}a 69, PL-00-681 Warszawa, Poland}

\author{Rafa{\l} Demkowicz-Dobrza{\'n}ski}
\affiliation{Institute of Theoretical Physics, University of Warsaw, ul. Ho\.{z}a 69, PL-00-681 Warszawa, Poland}

\begin{abstract}
We find the optimal scheme for quantum phase estimation in the presence of loss when no a priori knowledge on the estimated phase
is available. We prove analytically an explicit lower bound on estimation uncertainty, which
shows that, as a function of number of probes, quantum precision enhancement amounts at
most to a constant factor improvement over classical strategies.
\end{abstract}

\pacs{03.65.Ta, 06.20.Dk, 42.50.St}

\maketitle
\section{Introduction}
Owing to highly promising predictions of the theory of precise quantum
measurements and parameter estimation, as well as significant progress in quantum state engineering,
the task of phase shift determination
has recently been readdressed both theoretically and experimentally
\cite{Bollinger1996, Dowling1998, Berry2000, Giovannetti2004, Mitchell2004, Walther2004, Eisenberg2005, Higgins2007, Nagata2007, Kacprowicz2010,
Chwedenczuk2010}.
In classical
systems the precision of the estimated phase scales with the amount
of available resources as $1/\sqrt{N},$ the so called \emph{Standard
Quantum Limit} (SQL) or more commonly the {}``shot noise''. Traditionally,
$N$ denotes the number of independent measuring probes, repetitions
or copies of a system. The potential precision boost offered by quantum mechanics stems from the
possibility of preparing $N$ copies of a system in a highly entangled state,
particularly sensitive to the variations of the estimated parameter \cite{Bollinger1996, Dowling1998, Berry2000}.
In ideal scenarios, these states yield phase estimation precision
which scales as $1/N$ and is referred to as the \emph{Heisenberg Limit} (HL).

Environmentally induced decoherence, however, significantly affects
the performance of entanglement based quantum strategies
\cite{Paris1995, Huelga1997, Rubin2007, Olivares2007, Gilbert2008, Huver2008, Dorner2008, Demkowicz2009a, Banaszek2009, Ono2010}
with photon loss being its most relevant source in optical implementations.
The need to balance the phase sensitivity and robustness against losses results in
states performing better than SQL yet falling short of HL \cite{Dorner2008,Demkowicz2009a}.
Other approaches, trying to mimic the quantum enhanced strategies using multiple-pass
technique \cite{Higgins2007} are even more susceptible to losses and cannot compete with
the optimally designed entangled states \cite{Demkowicz2010}.
Despite the quantitative improvement of precision offered by quantum states in the presence of loss,
it has remained an unsolved problem
whether in the asymptotic regime $N \rightarrow \infty$
quantum states offer better than SQL scaling, i.e., $c/N^\alpha$ with $\alpha>1/2$.

In this paper we solve the problem of optimal phase estimation in the presence of loss with no a priori knowledge,
and prove analytically that even for arbitrarily small loss, quantum enhancement
does not offer better than $c/\sqrt{N}$ scaling for $N \rightarrow \infty$, and the only gain over classical strategies
is a smaller multiplicative constant $c$. It should be emphasized that the proof contains the most general description of a quantum
measurement, hence its conclusions are valid also for adaptive schemes (see Appendix \ref{sec:adaptmeas}), which are especially interesting from
a practical point of view \cite{Berry2009, Hentschel2010}.

\section{Model}
Two approaches to phase estimation are typically pursued. In the first,
\emph{local approach}, a measurement scheme is devised, which offers the highest sensitivity
to phase deviations from an a priori known value, $\varphi=\varphi_{0}$.
This is achieved by finding a strategy that maximizes the \emph{quantum
Fisher information}, $F_{Q}$, which defines the
lower bound on the precision of the estimated phase through $\delta\varphi\ge 1/\sqrt{F_{Q}}$
\cite{Helstrom1976,Holevo1982,Braunstein1994,Braunstein1996}. The optimal states have been
found both for lossless \cite{Bollinger1996, Dowling1998} (the so called N00N states) and more realistic lossy scenarios \cite{Dorner2008,Demkowicz2009a}.

The second approach, which we will pursue in this paper and refer to as the \emph{global approach},
assumes \emph{no a priori knowledge} about the phase, so that $\varphi$
is equiprobably distributed over the $[0,2\pi)$ region.
\begin{figure}[t]
\includegraphics[width=0.45 \textwidth]{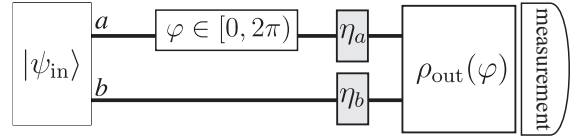}
\caption{Phase estimation setup. Channel $a$
  acquires a phase $\varphi$ relative to channel $b$. Losses are modeled by
  two beam splitters with power transmissions $\eta_a$, $\eta_b$.}
\label{fig:setup}
\end{figure}
We consider a general pure $N$ photon two-mode state \cite{Note1}
\begin{equation}
\ket{\psi_{\t{in}}} = \sum_{n=0}^N \alpha_n \ket{n ,N-n},
\end{equation}
which is
fed into an interferometer with a relative phase delay $\varphi$ (see Fig.~\ref{fig:setup}).
Apart from acquiring the phase via the unitary $U_\varphi = \mbox{e}^{-i \varphi a^\dagger a}$,
the state experiences losses modeled by two beam splitters with
 power transmissions $\eta_a$ and $\eta_b$ \cite{Note2}
 .
 The output state then takes the form
$\rho_{\t{out}}(\varphi)=U_\varphi \rho_{\t{out}} U^\dagger_{\varphi}$, where
\begin{equation}
\label{eq:rhoout}
\rho_{\t{out}}= \sum_{l_a=0}^N \sum_{l_b=0}^{N-l_a} \ket{\phi^{l_a,l_b}} \bra{\phi^{l_a,l_b}},
\end{equation}
with subnormalized conditional states corresponding to $l_a$ and $l_b$ photons lost in arms $a$ and $b$ respectively
\begin{equation}
\ket{\phi^{l_a,l_b}} = \sum_{n=l_a}^{N-l_b} \alpha_n \beta^{l_a,l_b}_{n} \ket{n-l_a,N - n- l_b}
\end{equation}
where
\begin{equation}
\beta_n^{l_a,l_b} = \sqrt{B_{l_a}^{n}(\eta_a) B_{l_b}^{N-n}(\eta_b)}, \ B_l^n(\eta)=\binom{n}{l}(1-\eta)^l \eta^{n-l}.
\end{equation}
Keeping the reasoning most general, the information about $\varphi$ is extracted via a measurement on $\rho_{\t{out}}(\varphi)$ described
by a Positive Operator Valued Measure (POVM), $\left\{ M_{r}\right\}$, $\sum_r M_{r}=\openone$.
The outcome $r$ is observed with probability
$p\left(r|\varphi\right)=\mbox{Tr}\left\{ \rho_{\t{out}}(\varphi)M_{r}\right\}$,
and the estimated phase inferred from it is defined
by an estimator $\tilde{\varphi}\left(r\right)$.
Optimization procedure with respect to a given cost function $C(\varphi,\tilde{\varphi})$ amounts to finding the state $\ket{\psi}$,
the measurement $\left\{ M_{r}\right\}$, and the estimator  $\tilde{\varphi}\left(r\right)$ that minimize
the cost function averaged over a flat a priori phase distribution
\begin{equation}
\left\langle C \right\rangle =
\int\frac{\mbox{d}\varphi}{2 \pi} \sum_{r}p\left(r|\varphi\right)C\left(\varphi,\tilde{\varphi}\left(r\right)\right).
\end{equation}
Let $C(\varphi, \tilde{\varphi})= C(\varphi - \tilde{\varphi}) = \sum_{n=-\infty}^{\infty} c_n e^{i n (\varphi-\tilde{\varphi})}$,
be an arbitrary real symmetric cost function ($c_n=c_{-n} \leq 0$ for $n \neq 0$) respecting the cyclic
nature of $\varphi$ \cite{Holevo1982, Chiribella2005}.

\section{Optimization}
Thanks to the flat a priori phase distribution, the problem enjoys a symmetry with respect to an arbitrary phase shift $U_\varphi$.
The search for the optimal measurement strategy may be restricted to the class of covariant POVM  $\{ M_{\tilde{\varphi}}\}$
\cite{Holevo1982, Chiribella2005, Bartlett2007} parameterized by a continuous parameter $\tilde{\varphi}$:
$M_{\tilde{\varphi}} = U_{\tilde{\varphi}} \Xi U^\dagger_{\tilde{\varphi}}$,
where $\Xi$ is a positive semi-definite operator satisfying the POVM completeness constraint
  $\int\frac{\t{d}\tilde{\varphi}}{2\pi}U_\varphi \Xi U^\dagger_\varphi= \openone$.
With the above substitution, the average cost function simplifies to
\begin{equation}
\left\langle C\right\rangle =\int\frac{\mbox{d}\varphi}{2\pi}\mbox{Tr}\left\{ \rho_{\t{out}}\left(\varphi\right)\Xi\right\}
C(\varphi)  \label{eq:average_gain}
\end{equation}
and $\left\langle C\right\rangle$ has to be minimized only over the
choice of the input state $\ket{\psi}_{\t{in}}$ and the seed operator $\Xi$.

In order to find the optimal $\Xi$, one can rewrite Eq.~(\ref{eq:rhoout})
in the form
$\rho_{\t{out}}=\bigoplus_{N^\prime=0}^N \rho^{N^\prime}_{\t{out}} $,
with $\rho^{N^\prime}_{\t{out}} = \sum_{l_a=0}^{N-N^\prime} \ket{\phi^{l_a,N - N^\prime -l_a}} \bra{\phi^{l_a,N- N^\prime -l_a}}$,
which reveals the block structure with respect to the total number of surviving photons $N^\prime$.
Therefore, without loss of generality, we may impose an analogous block structure on the seed operator $\Xi=\bigoplus_{N^\prime =0}^{N} \Xi^{N^\prime}$.
Physically, such a block structure implies that a non-demolition photon number measurement had been performed at the output, before
any further phase measurements have taken place.
Following the reasoning presented in \cite{Holevo1982,Chiribella2005} it can be shown
that without loosing optimality, the input state parameters $\alpha_n$ can be chosen real, in which case the optimal seed operator  $\Xi^{N^\prime}_{\t{opt}}=\left|e_{N'}\right\rangle \left\langle e_{N'}\right|$,
where $\left|e_{N'}\right\rangle =\sum_{n=0}^{N'}\left|n,N'-n\right\rangle $ (see Appendix \ref{sec:optmeasure}).

In what follows we choose the cost function $C({\varphi - \tilde{\varphi}}) = 4 \sin^2\frac{{\varphi-\tilde{\varphi}}}{2}$ ($c_0=2, c_1=c_{-1}=-1 $)
and denote its average by $\widetilde{\delta^2 \varphi}$, as
it is the simplest cost function approximating the variance for narrow distributions \cite{Berry2000}.

Performing the integration in Eq.~(\ref{eq:average_gain}) the average cost function reads:
\begin{equation}
\widetilde{\delta^2 \varphi}=2-\boldsymbol{\alpha}^\dagger \mathbf{A} \boldsymbol{\alpha},
\label{eq:matrix}
\end{equation}
where non-zero elements of the matrix $\mathbf{A}$ read:
\begin{align}
A_{n-1,n}  = A_{n,n-1}=\sum_{l_{a},l_{b}=0}^{n,N-n}
\beta^{l_a,l_b}_{n} \beta^{l_a,l_b}_{n-1}.
\end{align}
Hence, the minimal cost equals
$\widetilde{\delta^2\varphi}  = 2-\lambda_{\t{max}}$,
where $\lambda_{\t{max}}$ is the maximal eigenvalue of the matrix $\mathbf{A}$, and the corresponding eigenvector
provides the optimal input state parameters $\boldsymbol{\alpha}$.

\begin{figure}[t]
\includegraphics[width=1 \columnwidth]{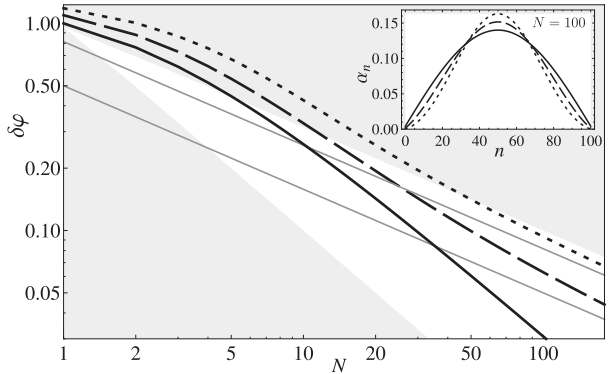}
\caption{Log-log plot of optimal phase estimation uncertainty as a function of number of photons used
for three different levels of loss (equal in both arms):
$\eta=1$ (solid), $\eta=0.8$ (dashed), $\eta=0.6$ (dotted). White area in the middle of the picture corresponds to
$1/N < \delta \varphi < 1/\sqrt{N}$. Gray lines represent asymptotic bounds given by Eq.~(\ref{eq:quantumequal}) for $\eta=0.8$, $\eta=0.6$. The inset depicts the structure of the optimal states for the three levels of loss for $N=100$.}
\label{fig:loglog}
\end{figure}
\subsection{Numerical solution}
Numerical results of the above eigenvalue problem are presented
in Fig.~\ref{fig:loglog}. Black lines depict phase estimation uncertainty
$\delta \varphi$ of the optimal quantum strategy plotted as a function of $N$ for  $\eta_a = \eta_b \in \{0.6,0.8,1\}$.
In the absence of loss the optimal quantum curve tends to the Heisenberg scaling,
whereas, when losses are present, it flattens significantly with increasing $N$.
The inset depicts the form of the optimal state. With increasing degree of loss
the distribution of $\alpha_n$ for the optimal state becomes more peaked as compared with the lossless case
$\alpha_n=\sqrt{\frac{2}{N+2}} \sin\left[\frac{\left(n+1\right)\pi}{N+2}\right]$ \cite{Berry2000}.
This behavior can be intuitively understood in a similar
fashion as in the local approach \cite{Dorner2008,Demkowicz2009a},
where the $N00N$ states with only two non-zero coefficients
$\alpha_0,\alpha_N$, are the most sensitive to the phase shift but
extremely vulnerable to loss. In the presence of loss, larger
weights need to be ascribed to intermediate coefficients, in order to preserve
quantum superposition even after some photons are lost. The same
effect of increasing weights of intermediate coefficients at the expense of marginal ones is also present in the global approach.

\subsection{Asymptotic bounds}
We now move on to present the main result of the paper.
Numerical results presented above
and the ones obtained within the local approach
\cite{Dorner2008,Demkowicz2009a} indicate that in the presence of loss,
phase estimation uncertainty $\delta \varphi$
departs from the HL and asymptotically approaches $c/\sqrt{N}$.
Until now, however, an analytical proof of the above conjecture was missing.



Let us first derive an upper bound on the maximal eigenvalue $\lambda_{\t{max}}$
of matrix $\mathbf{A}$ in Eq.~(\ref{eq:matrix}).
Without loss of generality, we assume that $\eta_a \leq \eta_b$.
Clearly, setting $\eta_b=1$ can only improve our estimation---hence
$\lambda_{\t{max}}$ increases. For $\eta_a=\eta < 1$, $\eta_b=1$ the nonzero matrix elements read:
$A_{n,n-1}= \sum_{l=0}^n \sqrt{B^n_l(\eta) B^{n-1}_l(\eta)}$.

 Recall that for an arbitrary normalized vector $\boldsymbol{v}$, $\boldsymbol{v}^\dagger \mathbf{A} \boldsymbol{v}\leq
 \lambda_{\t{max}}$. Let $\boldsymbol{\alpha}$ be
the eigenvector corresponding to $\lambda_{\t{max}}$:
$\boldsymbol{\alpha}^\dagger \mathbf{A} \boldsymbol{\alpha} = \lambda_{\t{max}}$.
The fact that all matrix elements of $\mathbf{A}$ are non-negative,
implies $\forall_n \alpha_n \geq 0$.

Let us now define a matrix $\mathbf{A}^\prime$,
such that all nonzero entries of $\mathbf{A}$ are replaced by
the maximum matrix element $A^\uparrow =\max_{n}
\left\{ A_{n,n-1}\right\} = A_{N,N-1} $.
Since $\alpha_n \geq 0$ and $A^\prime_{n,m} \geq A_{n,m} \geq 0$
we can write:
\begin{equation}
\lambda_{\t{max}} = \boldsymbol{\alpha}^\dagger \mathbf{A} \boldsymbol{\alpha} \leq
\boldsymbol{\alpha}^\dagger \mathbf{A}^\prime \boldsymbol{\alpha} \leq \lambda^\prime_{\t{max}},
\end{equation}
where $\lambda^\prime_{\t{max}}$ is the maximal eigenvalue of $\mathbf{A}^\prime$. $\lambda^\prime_{\t{max}}$ can be found analytically by noting the following recurrence
relation for the characteristic polynomial of $\mathbf{A^\prime}$:
$\t{det}\mathbf{\Lambda}_{n+1} = -\lambda \det{\mathbf{\Lambda}_n} -
A^{\uparrow 2} \det{\mathbf{\Lambda}_{n-1}}$, where
$\mathbf{\Lambda} = \mathbf{A}^\prime - \lambda \openone$,
while $\mathbf{\Lambda}_n$ are $(n+1) \times (n+1) $ submatrices of $\mathbf{\Lambda}$. The solution of the recurrence relation
reads
$\mbox{det}\left(\mathbf{\Lambda}\right)=
D_{N+1}(-\lambda,A^{\uparrow 2} ),$
where $D_{n}(x,a^{2})=a^{n}\frac{\mbox{sin}[(n+1)\mbox{arccos}(\frac{x}{2a})]}{\mbox{sin}[\mbox{arccos}(\frac{-x}{2a})]}$
is the Dickson polynomial \cite{Neusel2001} of the $n$th order.
The largest eigenvalue corresponds to the largest root of $\det\left(\mathbf{\Lambda}\right)$,
$\lambda^\prime_{\t{max}}=2 A^\uparrow \mbox{cos}\left[\frac{\pi}{\left(N+2\right)}\right]$.

We can finally write explicitly the lower bound on the variance:
\begin{equation}
\widetilde{\delta^2\varphi} \geq 2 \left[1 -   \cos{\left(\frac{\pi}{N+2}\right)} \sum_{l=0}^N
\sqrt{B_l^N(\eta) B^{N-1}_l(\eta)} \right].
\end{equation}
Expanding the above formula in the limit $N \rightarrow \infty$ we get:
\begin{equation}
\label{eq:quantum}
\widetilde{\delta^2\varphi} \geq \frac{1 - \eta}{4 \eta N} + O\left( \frac{1}{N^2}\right),
\end{equation}
which proves that for $\eta<1$, $\delta \varphi$ scales as
$c/\sqrt{N}$.

A tighter bound can be analogously derived for the case $\eta_a=\eta_b=\eta$, by noting
that $\max_{n} \left\{ A_{n,n-1}\right\} = A_{\left\lceil\frac{N}{2}\right\rceil,\left\lceil\frac{N}{2}\right\rceil-1}$.
In the limit $N \rightarrow \infty$ we get:
\begin{equation}
\label{eq:quantumequal}
\widetilde{\delta^2\varphi} \geq \frac{1 - \eta}{\eta N} + O\left( \frac{1}{N^2}\right).
\end{equation}

\subsection{Optimal classical strategy}
For the sake of comparison, we also derive the optimal classical phase estimation strategy,
in which a coherent state with mean photon number $N$
is sent to an initial beam splitter of transmissivity $\tau_{\t{in}}$, whose output
feeds paths $a$ and $b$ of the interferometer.
We assume no additional external phase
reference, hence the state is effectively a mixture
 of terms with a different total photon number. The optimal seed POVM is $\bigoplus_{N^\prime=0}^\infty \Xi^{N^\prime}_{\t{opt}}$ yielding:
\begin{equation}
\widetilde{\delta^2 \varphi}
 =2 -\frac{2 \mathcal{B}\left[N \eta_{a}\tau_{\t{in}}\right]
\mathcal{B}\left[N \eta_{b}\left(1-\tau_{\t{in}}\right)\right]}{N\sqrt{\eta_{a}\tau_{\t{in}}\,\eta_{b}\left(1-\tau_{\t{in}}\right)}}.
\end{equation}
where $\mathcal{B}(x)=\mbox{e}^{-x}\sum_{n=0}^{\infty}\frac{x^{n}}{n!}\sqrt{n}$
is the Bell polynomial of order $1/2$.
For strong beams ($N \rightarrow \infty$)
up to the first order in $1/N$,
 $\widetilde{\delta^2 \varphi} \approx \left(\frac{1}{\tau_{\t{in}} \eta_a} + \frac{1}{(1-\tau_{\t{in}}) \eta_b}  \right)/4N $ and is minimized for the
 choice $\tau_{\t{in}} = 1/(1+\sqrt{\eta_a/\eta_b})$
 \begin{equation}
 \label{eq:class}
  \widetilde{\delta^2 \varphi} \approx \frac{1}{4N} \left( \frac{1}{\sqrt{\eta_a}} + \frac{1}{\sqrt{\eta_b}} \right)^2,
   \end{equation}
 which is exactly the same formula as for the optimal classical
 strategy in the local approach \cite{Demkowicz2009a}.


\section{Conclusions}
Results presented in the paper indicate that, while quantum enhanced protocols provide
quantitative boost in the estimation precision, the presence of loss unavoidably causes
the precision scaling to become classical in the limit of large number of resources $N$.
The asymptotic gain of quantum enhanced protocols amounts just to a smaller multiplicative constant $c$ in the scaling
law $c/\sqrt{N}$. Comparing Eq.~(\ref{eq:class}) (with $\eta_a=\eta$, $\eta_b=1$)
 with the bound given in Eq.~(\ref{eq:quantum}) we may conclude that asymptotically quantum
 enhanced protocols provide at most a factor of
 \begin{equation}
 \lim_{N \rightarrow \infty}\frac{\delta \varphi^{\t{classical}}}{\delta \varphi^{\t{quantum}}} \leq \sqrt{\frac{1+\sqrt{\eta}}{1 - \sqrt{\eta}}}
 \end{equation}
 decrease in the uncertainty of estimation. In the case $\eta_a=\eta_b=\eta$, using a tighter bound (\ref{eq:quantumequal})
 the above factor reads $1/\sqrt{1-\eta}$.
We conjecture that the fact that losses necessarily turn HL into $c/\sqrt{N}$ is a general feature of all quantum estimation problems, such as
estimation of direction, Cartesian frames etc.

\begin{acknowledgements}
We acknowledge many fruitful discussions with Konrad Banaszek. This research was supported by the European Commission under
the Integrating Project Q-ESSENCE and the Foundation for Polish Science under the TEAM program.
\end{acknowledgements}

After this work has been completed,
analogous conclusions have been presented within the complementary local approach \cite{Knysh2010}.

\appendix

\begin{widetext}

\section{\label{sec:optmeasure}Optimal measurement}
Substituting the output state $\rho_{\t{out}}=\bigoplus_{N^\prime=0}^N \rho^{N^\prime}_{\t{out}}$
and the seed operator $\Xi=\bigoplus_{N^\prime =0}^{N} \Xi^{N^\prime}$ to Eq.~(6), we get an explicit formula
for the average cost function:
\begin{equation}
\label{eq:explicitcost}
\begin{split}
\left\langle C\right\rangle=\sum_{N^\prime=0}^{N}\sum_{l_{a}=0}^{N-N^\prime}\sum_{n,m=l_{a}}^{N^\prime+l_{a}}
C_{nm}\beta_{n}^{l_{a},l_b}\beta_{m}^{l_{a},l_b}\alpha_{n}^*\alpha_{m}\Xi_{n-l_{a},m-l_{a}}^{N^\prime},
\end{split}
\end{equation}
where $l_b=N-N^\prime-l_a$, $C_{nm}=\int\frac{\t{d}\varphi}{2\pi}C(\varphi)\mbox{e}^{i\left(n-m\right)\varphi}$
and $\Xi_{n^\prime,m^\prime}^{N^\prime}= \bra{n^\prime, N^\prime-n^\prime} \Xi^{N^\prime} \ket{m^\prime, N^\prime-m^\prime}$.
The completeness constraint $\int\frac{\t{d}\tilde{\varphi}}{2\pi}U_\varphi \Xi U^\dagger_\varphi= \openone$ implies that
$\Xi^{N^\prime}_{n^\prime n^\prime}=1$. Therefore, if restricted to $m=n$ terms, the sum (\ref{eq:explicitcost}) reduces to a constant
$c_0 = C_{00}$. Changing the summation order we can rewrite Eq.~(\ref{eq:explicitcost}) as
\begin{equation}
\label{eq:explicitcost2}
\left\langle C\right\rangle - c_0=\sum_{\underset{n\ne m}{n,m=0}}^{N}\sum_{l_a=0}^{\min(n,m)}\sum_{l_{b}=0}^{N-\max(n,m)}
C_{nm}\beta_{n}^{l_{a},l_b}\beta_{m}^{l_{a},l_b}\alpha_{n}^*\alpha_{m}\Xi_{n-l_{a},m-l_{a}}^{N-l_a-l_b}.
\end{equation}
Now, as for all $n \neq m$ cost coefficients $C_{nm} \leq 0$, we get the following lower bound
on the average cost
\begin{eqnarray}
\left\langle C\right\rangle -c_{0} & \ge &\sum_{\underset{n\ne m}{n,m=0}}^{N}\sum_{l_a=0}^{\min(n,m)}
\sum_{l_{b}=0}^{N-\max(n,m)} C_{nm}\beta_{n}^{l_{a},l_{b}}\beta_{m}^{l_{a},l_{b}}
\left|\alpha^*_{n}\right|\left|\alpha_{m}\right|\left|\Xi_{n-l_{a},m-l_{a}}^{l_{a}+l_{b}}\right|\\
\label{eq:indistinguish}
 & \ge & \sum_{\underset{n\ne m}{n,m=0}}^{N}\sum_{l_a=0}^{\min(n,m)}
 \sum_{l_{b}=0}^{N-\max(n,m)}C_{nm}\beta_{n}^{l_{a},l_{b}}\beta_{m}^{l_{a},l_{b}}\left|\alpha_{n}^*\right|\left|\alpha_{m}\right|.
 \end{eqnarray}
The first inequality is saturated by choosing input state's and seed operator's
coefficients to be real. The second inequality follows from
$\Xi_{n^\prime,m^\prime}^{N^\prime}\le\sqrt{\Xi_{m^\prime,m^\prime}^{N^\prime}\Xi_{n^\prime,n^\prime}^{N^\prime}}=1$,
which is a consequence of positive semi-defniteness of $\Xi^{N^\prime}$
and  the completeness constraint.
Both inequalities are saturated for $\Xi^{N^\prime}_{\t{opt}}=\left|e_{N'}\right\rangle \left\langle e_{N'}\right|$,
where $\left|e_{N'}\right\rangle =\sum_{n=0}^{N'}\left|n,N'-n\right\rangle$. This proves the optimality of the measurement
considered in the paper.

\section{\label{sec:distin}Distinguishability of photons}
If photons traveling through the interferometer are distinguishable, e.g. they are prepared in different time bins,
the dimension of the Hilbert space needed to describe the state of $N$ photons is $2^N$, as opposed to $N+1$
for the indistinguishable case. In fact, the indistinguishable case may be
considered as a restriction of the former space to its fully symmetric subspace. We prove below that considering distinguishable
photons is of no use, since the optimality can always be attained within the class of states belonging to the fully symmetric (bosonic) subspace.
Let
\begin{equation}
\ket{\psi_N} =\sum_{\mathbf{n}=\boldsymbol{0}^N}^{\boldsymbol{1}^N}\alpha_{\boldsymbol{n}}\left|\boldsymbol{n}\right\rangle,
\end{equation}
be a general state of $N$ distinguishable photons traveling through the interferometer,
where the sum runs over all $N$-bit sequences $\boldsymbol{n}$,  with $\left|\boldsymbol{n}\right\rangle =\left|n_{1}\right\rangle \otimes\dots\left|n_{N}\right\rangle $, where  $\left|n_{i}\right\rangle =\ket{1}$ ($\ket{0}$)  
denotes a photon in the $i$th time bin, propagating in the $a(b)$
arm of the interferometer respectively.

Taking loss into account, we additionally need to track the time slots in which photons were lost.
We define
a binary string $\boldsymbol{l}_{a}=l_{a,1}l_{a,2}\dots l_{a,N}$ with $1$s
representing the time bins in which photon was lost in arm $a$ and
similarly $\boldsymbol{l}_{b}$ for the arm $b$.
The general seed operator has a block diagonal structure with respect to different patterns
of surviving photons:
$\Xi = \bigoplus_{\boldsymbol{N}^\prime=\boldsymbol{0}^N}^{\boldsymbol{1}^N} \Xi^{\boldsymbol{N}^\prime}$, where
1s in the binary string $\boldsymbol{N}^\prime$ denote the time bins in which photons were successfully transmitted.
Formally, using bitwise subtraction, we can write $\boldsymbol{N}^\prime = \boldsymbol{1} - \boldsymbol{l}_a - \boldsymbol{l}_b$.
Written in a basis
$\Xi^{\boldsymbol{N}^\prime} = \sum_{\boldsymbol{n}^\prime, \boldsymbol{m^\prime} = \boldsymbol{0}^{N^\prime}}^{\boldsymbol{1}^{N^\prime}}
\Xi^{\boldsymbol{N}^\prime}_{\boldsymbol{n}^\prime,\boldsymbol{m}^\prime} \ket{\boldsymbol{n}^\prime} \bra{\boldsymbol{m}^\prime}$,
in which $\boldsymbol{n^\prime}$ stands for a string with $N^\prime$ bits placed at positions corresponding to 1s in $\boldsymbol{N}^\prime$
with complementary positions left empty (neither $0$ nor $1$).
In order to simplify the notation, for any binary sequence $\boldsymbol{x}$, we denote by $x=|\boldsymbol{x}|$ the number of 1s in the sequence.
Moreover, we use a notation $\boldsymbol{x}\setminus \boldsymbol{y}$ for a binary string $\boldsymbol{x}$ with empty entries at positions
corresponding to 1s in $\boldsymbol{y}$.

Adapting Eq.~(\ref{eq:explicitcost2}) to the distinguishable photon case, we get:
\begin{equation}
\left\langle C\right\rangle -c_{0}  =  \sum_{\underset{n\ne m}{\boldsymbol{n},\boldsymbol{m}=\boldsymbol{0}}}^{\boldsymbol{1}}
\sum_{\boldsymbol{l}_{a}=\boldsymbol{0}}^{\min(\boldsymbol{n},\boldsymbol{m})}
\sum_{\boldsymbol{l}_{b}=\boldsymbol{0}}^{\boldsymbol{1}-\max(\boldsymbol{n},\boldsymbol{m})}
C_{nm}
\gamma_n^{l_a,l_b} \gamma_m^{l_a l_b}
\alpha_{\boldsymbol{n}}^*\alpha_{\boldsymbol{m}}
\Xi_{\boldsymbol{n}\setminus (\boldsymbol{l}_a+\boldsymbol{l}_b),\boldsymbol{m}\setminus (\boldsymbol{l}_a+\boldsymbol{l}_b)}^{\boldsymbol{1} - (\boldsymbol{l}_a+\boldsymbol{l}_b)}
\end{equation}
where $\min$, $\max$ should be understood as bitwise operations, $\gamma_n^{l_a,l_b} =\sqrt{(1-\eta_a)^{l_a}\eta_a^{n-l_a} (1-\eta_b)^{l_b}\eta_b^{N-n-l_b} }$
and for simplicity we have put $\boldsymbol{0}=\boldsymbol{0}^{N}$, $\boldsymbol{1}=\boldsymbol{1}^{N}$.

We now split the sums over $\boldsymbol{l}_i$ into sum over $l_i$ (number of 1s in $\boldsymbol{l}_i)$
and the sum over permutation of 1s within $\boldsymbol{l}_i$. We proceed analogously
for summations over $\boldsymbol{n}$ ($\boldsymbol{m}$) obtaining
\begin{eqnarray}
\left\langle C\right\rangle -c_{0} & = &
\sum_{\underset{n\ne m}{n,m=0}}^{N}
\sum_{l_{a}=0}^{\min(n,m)}
\sum_{l_{b}=0}^{N-\max(n,m)}C_{nm}
\gamma_{n}^{l_{a},l_{b}}\gamma_{m}^{l_{a}l_{b}}\\
&  & \sum_{\underset{|\boldsymbol{n}|=n}
{\boldsymbol{n}=\boldsymbol{0}}}^{\boldsymbol{1}}
\sum_{\underset{|\boldsymbol{m}|=m}
{\boldsymbol{m}=\boldsymbol{0}}}^{\boldsymbol{1}}
\alpha_{\boldsymbol{n}}^{*}
\alpha_{\boldsymbol{m}}
\sum_{\underset{|\boldsymbol{l}_{a}|=l_{a}}
{\boldsymbol{l}_{a}=\boldsymbol{0}}}^{\min(\boldsymbol{n},\boldsymbol{m})}
\sum_{\underset{|\boldsymbol{l}_{b}|=l_{b}}{\boldsymbol{l}_{b}=\boldsymbol{0}}}^{\boldsymbol{1}-
\max(\boldsymbol{n},\boldsymbol{m})}\Xi_{\boldsymbol{n}\setminus(\boldsymbol{l}_{a}+\boldsymbol{l}_{b}),\boldsymbol{m}\setminus(\boldsymbol{l}_{a}+\boldsymbol{l}_{b})}^{\boldsymbol{1}-(\boldsymbol{l}_{a}+\boldsymbol{l}_{b})}\nonumber \label{eq:fullloss}
\end{eqnarray}
In order to proceed further let us for the moment specialize to lossless case $\eta_a=\eta_b=1$, where the above formula simplifies to:
\begin{equation}
\left\langle C\right\rangle -c_{0}   =
\sum_{\underset{n\ne m}{n,m=0}}^{N}
C_{nm}
\sum_{\underset{|\boldsymbol{n}|=n}{\boldsymbol{n}=\boldsymbol{0}}}^{\boldsymbol{1}}
\sum_{\underset{|\boldsymbol{m}|=m}{\boldsymbol{m}=\boldsymbol{0}}}^{\boldsymbol{1}}
\alpha_{\boldsymbol{n}}^*\alpha_{\boldsymbol{m}}
\Xi_{\boldsymbol{n},\boldsymbol{m}}^{\boldsymbol{1}}.
\end{equation}
$\Xi$ needs to be a positive semi-definite operator, and by completeness constraint $\Xi_{\boldsymbol{m},\boldsymbol{n}}=
\delta_{\boldsymbol{m},\boldsymbol{n}}$,
whenever $n=m$. Since diagonal blocks of $\Xi$ (corresponding to $n=m$) are proportional to identity, it implies that none
of the off-diagonal blocks of $\Xi$ (corresponding to $n\neq m$) can have a singular value larger than $1$.
This can be proven as follows.
Let us assume that for certain block $(m,n)$ ($n\neq m$), the largest singular value $\lambda >1$, and
let $\ket{\boldsymbol{v}_m}$, $\ket{\boldsymbol{w}_n}$ be the normalized left and right singular vectors corresponding to singular value $\lambda$,
$|\boldsymbol{v}_m|=m$,
$|\boldsymbol{w}_n|=n$.
Defining $\ket{\boldsymbol{z}}=\ket{\boldsymbol{v}_m}-\ket{\boldsymbol{w}_n}$, we calculate
\begin{equation}
\bra{\boldsymbol{z}} \Xi \ket{\boldsymbol{z}} = \bra{\boldsymbol{v}_n} \Xi \ket{\boldsymbol{v}_n} + \bra{\boldsymbol{w}_m} \Xi \ket{\boldsymbol{w}_m}
- 2 \Re \bra{\boldsymbol{v}_n} \Xi \ket{\boldsymbol{w}_m} = 2(1-\lambda) < 0,
\end{equation}
which contradicts the positivity semi-definiteness of $\Xi$.
Because all singular values of any (n,m) block of $\Xi$ are smaller than one, the following inequality holds:
$\sum_{\underset{|\boldsymbol{n}|=n}{\boldsymbol{n}=\boldsymbol{0}}}^{\boldsymbol{1}}
\sum_{\underset{|\boldsymbol{m}|=m}{\boldsymbol{m}=\boldsymbol{0}}}^{\boldsymbol{1}}
\alpha_{\boldsymbol{n}}^*\alpha_{\boldsymbol{m}}
\Xi_{\boldsymbol{n},\boldsymbol{m}} \leq \alpha_n^* \alpha_m$,
 $\alpha_n = \sqrt{\sum_{\underset{|\boldsymbol{n}|=n}{\boldsymbol{n}=\boldsymbol{0}}}^{\boldsymbol{1}} |\alpha_{\boldsymbol{n}}|^2}$.
This leads to a bound on the cost function in the lossless case
\begin{equation}
\left\langle C\right\rangle -c_{0}   \geq
\sum_{\underset{n\ne m}{n,m=0}}^{N} C_{nm} \alpha_n \alpha_m^*,
\end{equation}
proving that one can achieve optimality restricting oneself to indistinguishable photons.

Returning to Eq.~(\ref{eq:fullloss}), we see that we can apply a similar argumentation making use of positive semi-definiteness of
$\Xi^{(l_a,l_b)}_{\boldsymbol{m},\boldsymbol{n}}=\sum_{\underset{|\boldsymbol{l}_a|=l_a}{\boldsymbol{l}_a=\boldsymbol{0}}}^{\min(\boldsymbol{n},\boldsymbol{m})}
\sum_{\underset{|\boldsymbol{l}_b|=l_b}{\boldsymbol{l}_b=\boldsymbol{0}}}^{\boldsymbol{1} - \max(\boldsymbol{n},\boldsymbol{m})}
\Xi_{\boldsymbol{m}\setminus (\boldsymbol{l}_a+\boldsymbol{l}_b),\boldsymbol{n}\setminus (\boldsymbol{l}_a+\boldsymbol{l}_b)}^{\boldsymbol{1} - (\boldsymbol{l}_a+\boldsymbol{l}_b)}$
operator. We notice that the completeness constraint again implies a block structure of
 $\Xi^{(l_a,l_b)}$ with respect to $m = |\boldsymbol{m}|, n = |\boldsymbol{n}| $, with diagonal elements of diagonal blocks $(n,n)$
 being now $\sum_{\underset{|\boldsymbol{l}_a|=l_a}{\boldsymbol{l}_a=\boldsymbol{0}}}^{\boldsymbol{n}}
\sum_{\underset{|\boldsymbol{l}_b|=l_b}{\boldsymbol{l}_b=\boldsymbol{0}}}^{\boldsymbol{1} - \boldsymbol{n}} 1 =
\binom{n}{l_a} \binom{N-n}{l_b}$. This implies that the maximum singular value of any (m,n) block of $\Xi^{(l_a,l_b)}$
is constrained by $\binom{\min(n,m)}{l_a} \binom{N-\max(n,m)}{l_b}$. As a result, we obtain the following bound:
\begin{equation}
\begin{split}
\label{eq:fullloss2}
\left\langle C\right\rangle -c_{0}   \geq
\sum_{\underset{n\ne m}{n,m=0}}^{N}
\sum_{l_{a}=0}^{\min(n,m)}
\sum_{l_{b}=0}^{N-\max(n,m)}
C_{nm}
\gamma_n^{l_a,l_b} \gamma_m^{l_a l_b}
\binom{\min(n,m)}{l_a} \binom{N-\max(n,m)}{l_b} |\alpha_n| |\alpha_m^*| \geq \\
\sum_{\underset{n\ne m}{n,m=0}}^{N}
\sum_{l_{a}=0}^{\min(n,m)}
\sum_{l_{b}=0}^{N-\max(n,m)}
C_{nm}
\gamma_n^{l_a,l_b} \gamma_m^{l_a l_b}
\sqrt{\binom{n}{l_a}\binom{m}{l_a} \binom{N-n}{l_b}\binom{N-m}{l_b}} |\alpha_n| |\alpha_m^*|
\end{split}
\end{equation}
.
Recalling that $\beta_{n}^{l_a,l_b} = \sqrt{\binom{n}{l_a}\binom{N-n}{l_b}} \gamma_{n}^{l_a,l_b}$,
it is evident that the above equation is identical to Eq.~(\ref{eq:indistinguish}) obtained for the indistinguishable case. this completes the proof that
the optimal estimation is indeed achievable using indistinguishable photons.

\section{\label{sec:adaptmeas}Adaptive measurement schemes}
Let us describe a general structure of adaptive measurement schemes performed on $N$ subsystems.
Let $\{ \Pi^{(1)}_{i_1}\}$ be a POVM performed on the first copy. Depending on the measurement result $i_1$
a POVM $\{ \Pi^{(2)}_{i_2}(i_1) \}$ is performed on the second copy. In general, a POVM performed on the $k$-th copy
$\{ \Pi^{(k)}_{i_k}(i_1,\dots,i_{k-1})\}$ depends on all previous measurement results. The adaptive measurement
mathematically corresponds to POVM:
\begin{equation}
\Pi_{\boldsymbol{i}} = \Pi_{i_1,\dots,i_N} = \Pi^{(1)}_{i_1} \otimes \dots \otimes \Pi^{(N)}_{i_N}(i_1,\dots,i_{N-1}),
\end{equation}
where $\Pi_{\boldsymbol{i}}$ can be treated as a single global POVM with measurement results indexed by $\boldsymbol{i}$.
This shows that, for distinguishable subsystems, optimization of estimation strategy over global POVMs
covers also the case of adaptive measurements.
Moreover, we have proved earlier in Appendix \ref{sec:distin}, that the optimal phase estimation can be realized using
indistinguishable subsystems. Therefore, the bounds derived in the paper, which assume a global POVM on indistinguishable photons, indeed hold also for all
adaptive measurement strategies.
\end{widetext}

\bibliographystyle{apsrev4-1}
%

\end{document}